\documentclass[3p,times,procedia]{elsarticle}
\usepackage{nupha_ecrc}

\volume{00}

\firstpage{1}

\journalname{Nuclear Physics A}

\runauth{}

\jid{nupha}

\jnltitlelogo{Nuclear Physics A}

\usepackage{amssymb}

\usepackage[figuresright]{rotating}

\begin{document}

\begin{frontmatter}



\dochead{XXVIth International Conference on Ultrarelativistic Nucleus-Nucleus Collisions\\ (Quark Matter 2017)}

\title{Lattice Calculations of Heavy Quark Potential at Finite Temperature}


\author{P. Petreczky$^1$ and J. Weber$^2$ (for TUMQCD Collaboration)}
\address{$^1$ Physics Department, Brookhaven National Laboratory, Upton, NY 11973, USA}
\address{$^2$ Physik Department, Technische Universit\"{a}t M\"{u}nchen, D-85748 Garching, Germany}

\begin{abstract}
We report on the lattice calculations of the heavy quark potential at $T>0$
in 2+1 flavor QCD at physical quark masses using the Highly Improved Staggered
Quark discretization. 
We study in detail the systematic effects in the determination of the real and imaginary
parts of the potential when using the moment method.
\end{abstract}




\end{frontmatter}


\section{Introduction}
\label{}
The problem of quarkonium properties at high temperatures received a lot
of theoretical attention since the famous paper by Matsui and Satz \cite{Matsui:1986dk}
(see e.g. Ref. \cite{Mocsy:2013syh} for a recent review). A lattice calculation of the quarkonium
spectral functions would solve the problem. However, the reconstruction of meson spectral
functions from a discrete set of data points in Euclidean time proved to be an extremely challenging task (see
e.g. Refs. \cite{Wetzorke:2001dk,Datta:2003ww}.) Another method of obtaining quarkonium
spectral functions relies on effective field theory approaches.
One effective field theory approach is called NRQCD and studies based on this approach have been
reported in Refs. \cite{Kim:2014iga,Aarts:2014cda}.
Another effective field theory approach for calculating quarkonium spectral functions is pNRQCD 
\cite{Burnier:2007qm,Brambilla:2008cx}. The basic ingredients of the latter approach are
the heavy quark anti-quark potentials.
In the weakly-coupled regime these potentials can be calculated perturbatively \cite{Brambilla:2008cx},
while in the strong-coupling regime a lattice calculation is needed \cite{Petreczky:2010tk}.
Attempts to calculate the heavy quark anti-quark potential on the lattice using a Bayesian
approach have been presented \cite{Burnier:2014ssa}. A different solution to the problem, which is 
based on using a simple Ansatz for the spectral function and fitting
the lattice data, has been presented in Ref. \cite{Bazavov:2014kva}. In this contribution we
will follow on the latter approach extending it using the method of moments of correlation functions.

\section{Moments method and the numerical results}

To obtain the potential on the lattice we study the correlation function of an infinitely heavy (static)
quark anti-quark ($Q \bar Q)$ pair in Euclidean time $\tau$. The quark and anti-quark are separated by the 
distance $r$ and we use Coulomb gauge to define the correlation functions. We also considered HYP smeared
Wilson loops and found very similar results. The $Q\bar Q$ correlation functions have been calculated on
$48^3 \times 12$ lattice generated by HotQCD collaboration \cite{Bazavov:2014pvz}.
In our study we used the lattice gauge couplings $\beta=10/g^2=7.03,~7.28,~7.373,~7.596$ and $7.825$
corresponding to temperatures $T=199,~251,~273,~333$ and $407$ MeV, respectively.
The correlation function has the following spectral representation for $\tau < 1/T$
\begin{equation}
W_r(\tau,T)=\int_{-\infty}^{\infty} d \omega \sigma_r(\omega,T) e^{-\omega \tau}
\end{equation}
The real part of the potential corresponds to the lowest peak position in $\sigma_r(\omega,T)$,
while the imaginary part of the potential corresponds to the width of the peak.
The moments of the correlation functions are defined as
\begin{equation}
m_1(\tau,r,T)=- \frac{\partial_{\tau} W_r(\tau,T)}{W_r(\tau,T)},~~m_n=-\partial_{\tau} m_{n-1}(\tau,r,T),~n>1.
\end{equation}
The first moment is just the so-called effective mass used in lattice calculations
at zero temperature. For large $\tau$ it approaches the energy of the ground state.
In some limiting cases the above moments are related to the 
moments of the spectral function around the peak. For example, in the case
of a Gaussian spectral function we have 
\begin{equation}
m_1(\tau)=\frac{ \int_{-\infty}^{\infty} d \omega \sigma(\omega,T) e^{-\omega \tau} }{ \int_{-\infty}^{\infty} d \omega \sigma(\omega,T) 
e^{-\omega \tau} } = \langle \omega \rangle \simeq {\rm Re} V-({\rm Im} V)^2 \tau,
\end{equation}
which for small ${\rm Im} V$ or $\tau$ is just the real part of the potential.
Similarly we find that $m_2 \simeq ({\rm Im} V)^2$. As we will see below the moments $m_n$ provide
an efficient way to summarize the information contained in the static $Q \bar Q$ correlator.

\begin{figure}
\includegraphics[width=7cm]{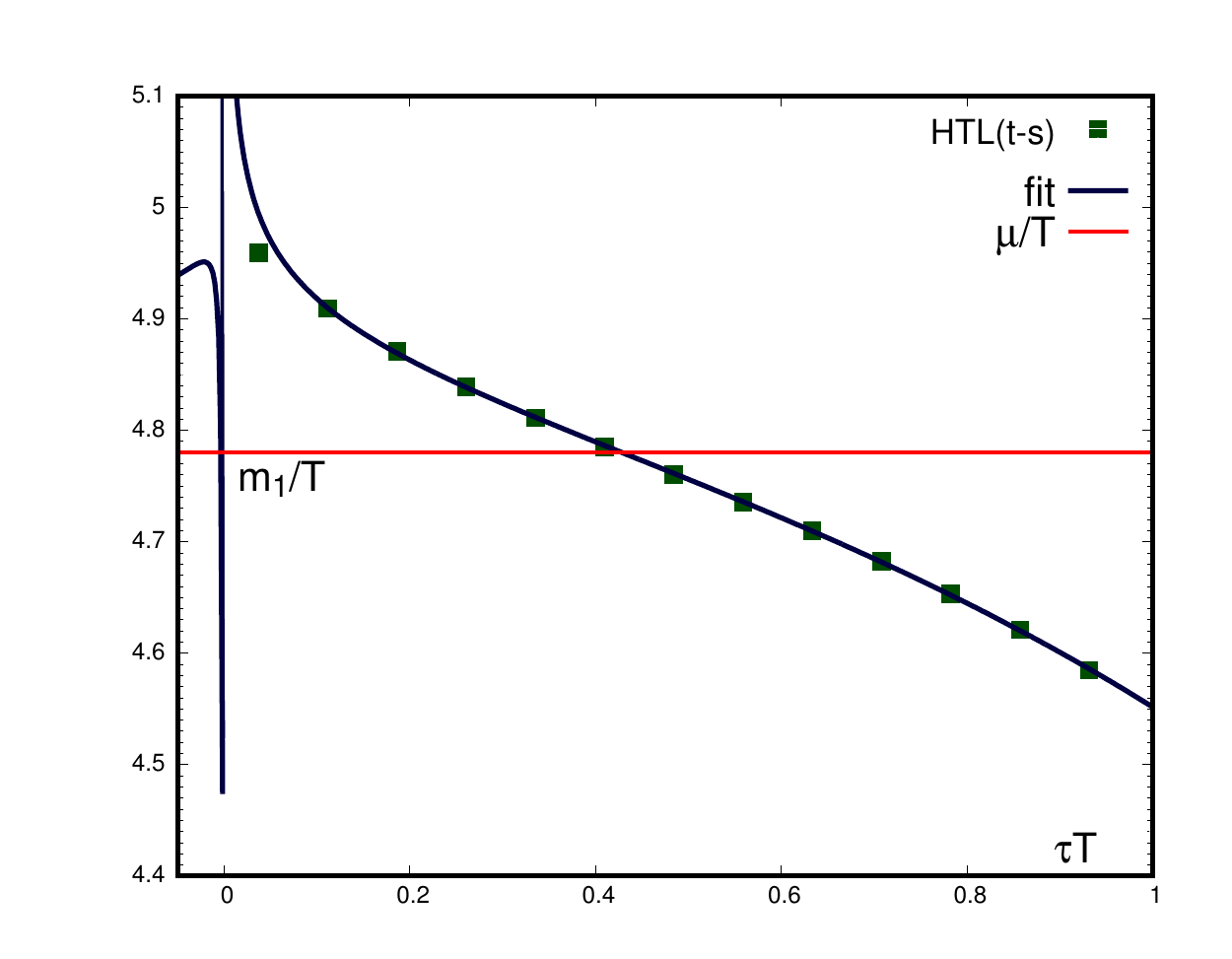}
\includegraphics[width=7cm]{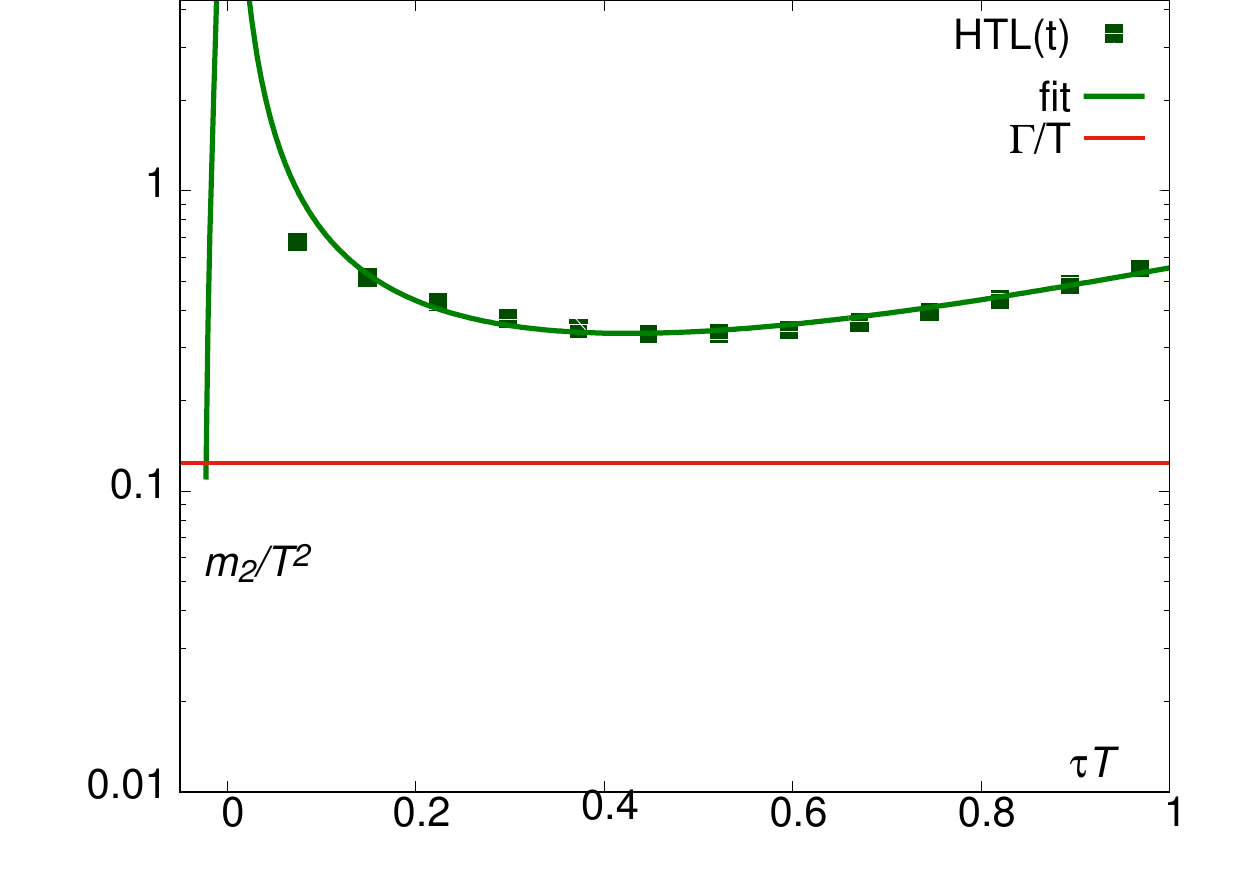}
\caption{The first and second moments of the Euclidean $Q \bar Q$ correlator
obtained in HTL perturbation theory and fitted with Lorentzian form.
The horizontal lines correspond to the values of $\mu$ and $\Gamma$ obtained from the fit.}
\label{fig:htl}
\end{figure}

\begin{figure}
\includegraphics[width=7.3cm]{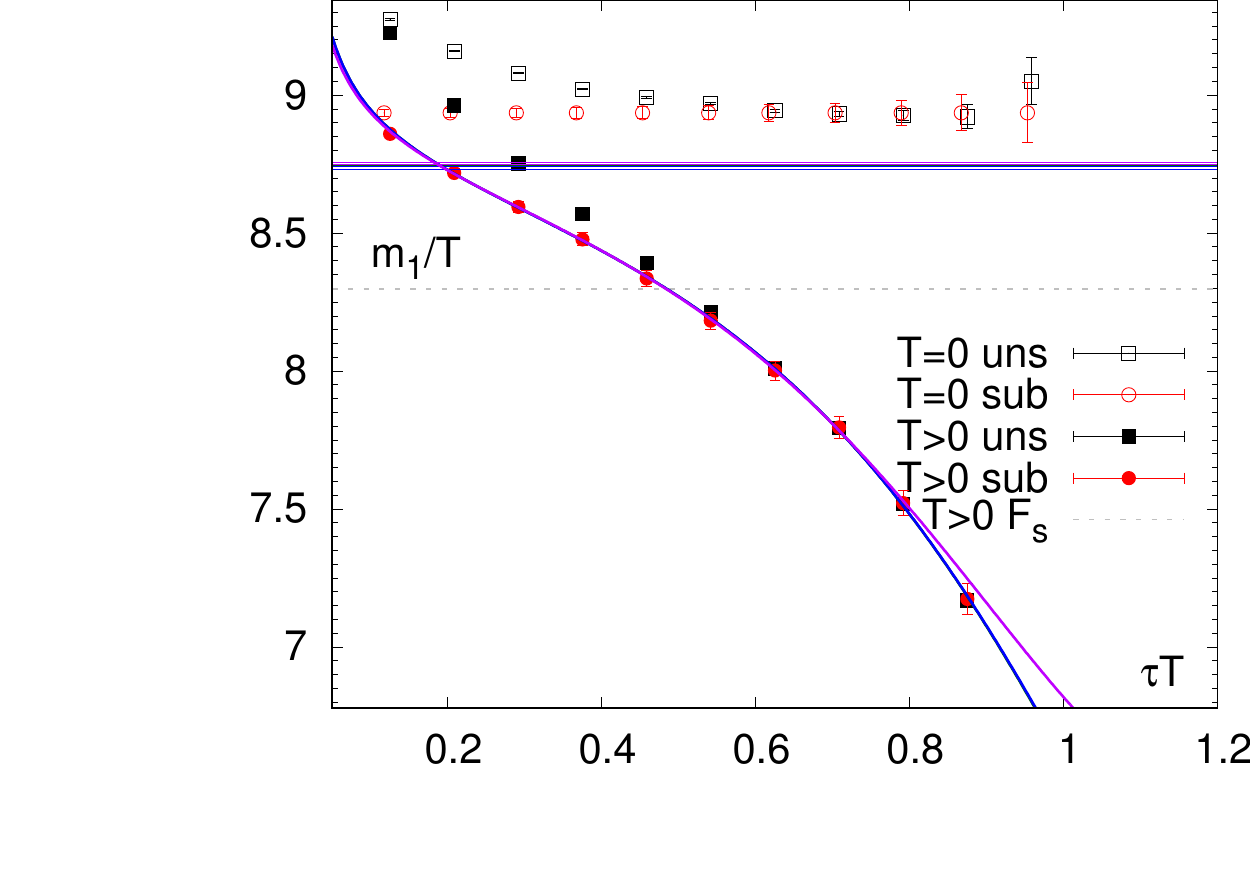}
\includegraphics[width=7.3cm]{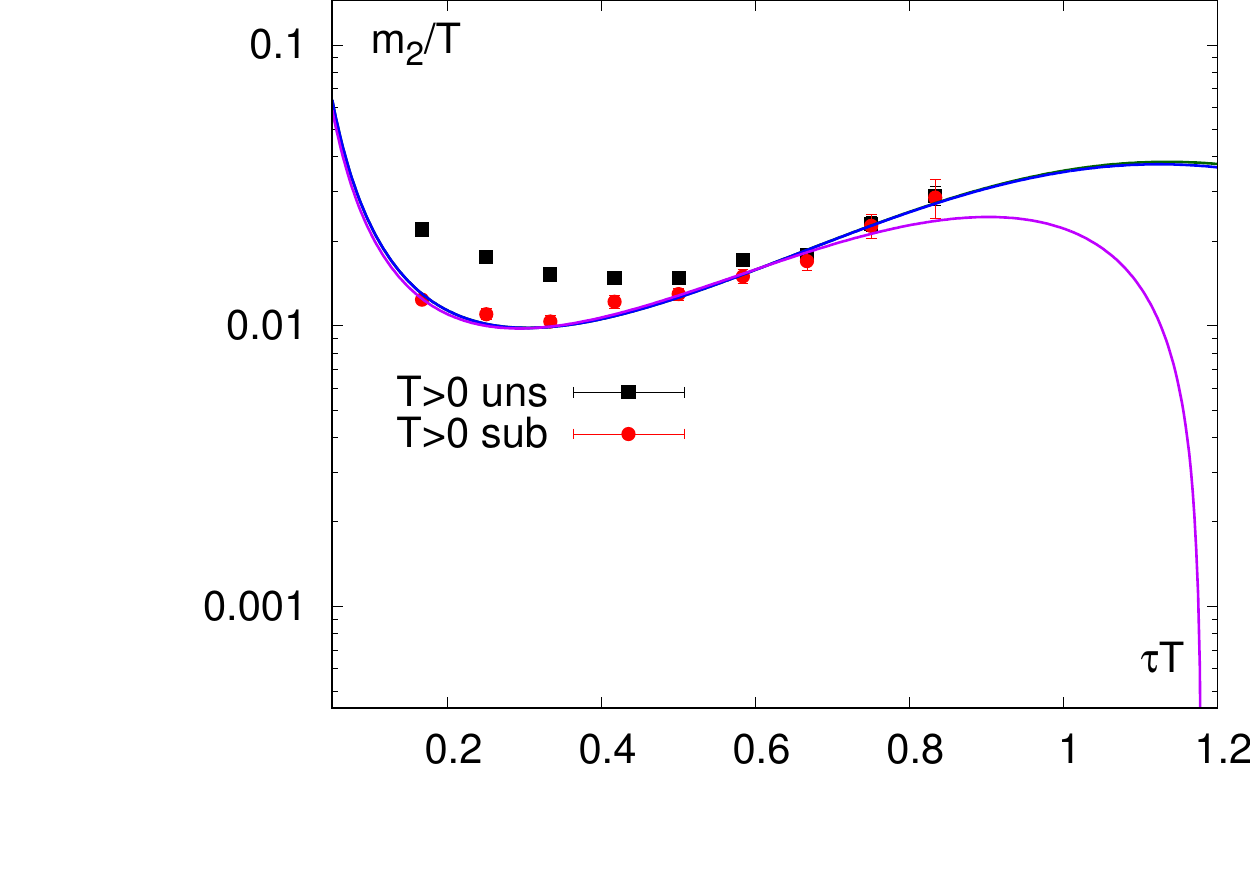}
\caption{The first (left) and the second (right) moments of
the correlators calculated for $\beta=7.825$ ($T=407$ MeV)
at zero and finite temperature.}
\label{fig:mom}
\end{figure}

On the lattice the moments are defined as
\begin{eqnarray}
&
m_1(\tau-\frac{a s}{2},r,T)=-\frac{1}{s}\ln\left(\frac{W_r(\tau-a s,T)}{W_r(\tau,T)}\right) \\
&
m_n(\tau-\frac{a s}{2},r,T)=\frac{1}{s}\left(m_{n-1}\left(\tau,r,T\right)-m_{n-1}(\tau-\frac{s}{a},r,T)\right),~n=2,3\dots, s=1,2 \dots
\end{eqnarray}
In our investigations we use a Lorentzian form of the spectral function:
\begin{equation}
\sigma_r(\omega,T)=\frac{1}{\pi} \frac{\Gamma}{(\omega-\mu)^2+\Gamma^2}\Theta(\omega-\eta),~
\mu={\rm Re} V(r,T),~\Gamma ={\rm Im} V(r,T)
\end{equation}
The Lorentzian form alone is not expected to describe the spectral function accurately away from the peak.
For this reason we included a regulator parameter $\eta$ that cuts off the low $\omega$ tail of the Lorentzian.
For large $\omega$ no regulator is needed as the exponential kernel in the Laplace transform cuts off the tail
of the Lorentzian.
To check the viability of this simple form we use the spectral functions calculated in HTL perturbation
theory \cite{Burnier:2013fca} and fit the corresponding Euclidean correlation function with the above Lorentzian form. The
results of these fits are shown in Fig. \ref{fig:htl}. Except for very small $\tau$ the fits can describe
the HTL perturbative results well, even-though 
the HTL spectral functions are not of Lorentzian shape \cite{Burnier:2013fca}. 
The input values of the peak position ($\mu/T=4.72$) and width ($\Gamma/T=0.122$) are reproduced well.
Thus, we conclude that a modified Lorentzian form can be used for  parametrizing the spectral functions
in our study.
Our numerical results for the first and second moments are shown in Fig.
\ref{fig:mom} for $T=407$ MeV and $rT=0.85$. The results for other temperature and
$r T$ values are similar.
In the figure we show the zero temperature results for $m_1$ as well. At zero temperature
$m_1$ approaches a plateau for $\tau T>0.6$, but this is not the case at finite temperature.
The decrease of $m_1$ at finite temperature at large $\tau$ is qualitatively similar to the one observed
in HTL perturbation theory and thus could be associated with the imaginary part of the potential.
Since at small $\tau$ the high $\omega$ part of the spectral function, which corresponds
to the excited states  is important we need to isolate the corresponding contribution in the correlator.
It is customary to parametrize the $\tau$ dependence of the correlator in terms of two exponentials.
The first exponential corresponds to the ground state, while the second exponential corresponds
to the higher excited states and can parametrize the high $\omega$ part of the spectral function.
We performed a double exponential fit of the zero temperature correlator and subtracted the 
contribution due to the second exponent. The value of $m_1$ calculated from the subtracted
correlator is also shown in Fig. \ref{fig:mom}. As one can see from the figure the zero temperature $m_1$ from
the subtracted correlator shows a plateau for all $\tau$ values. Since at very large $\omega$ we
do not expect the spectral function to be temperature dependent we perform the same subtraction
for the finite temperature correlator, which makes the value of $m_1$ smaller
for small $\tau$ but has no effect on its large $\tau$ behavior. 
The subtracted correlator then is fitted with the Lorentizan
form. We performed different fits, varying the fit interval in $\tau$ and
using different values of the $\eta$ parameter. The fits work very well
and the values of $\mu$ and $\Gamma$ are not very sensitive to the
choice of the fit interval and the value of $\eta$.
The results of the fits are also shown in Fig. \ref{fig:mom}. 
The horizontal lines in Fig. \ref{fig:mom} (left) show the results for $\mu$.
In addition we calculated $m_2$ from the subtracted correlator and the results are shown
in Fig. \ref{fig:mom} (right). At small $\tau$ the second moment decreases almost by
a factor of three as the result of subtraction. Our fits describe the subtracted $m_2$ well even-though the corresponding
data do not enter the fits. 

\begin{figure}
\includegraphics[width=7.3cm]{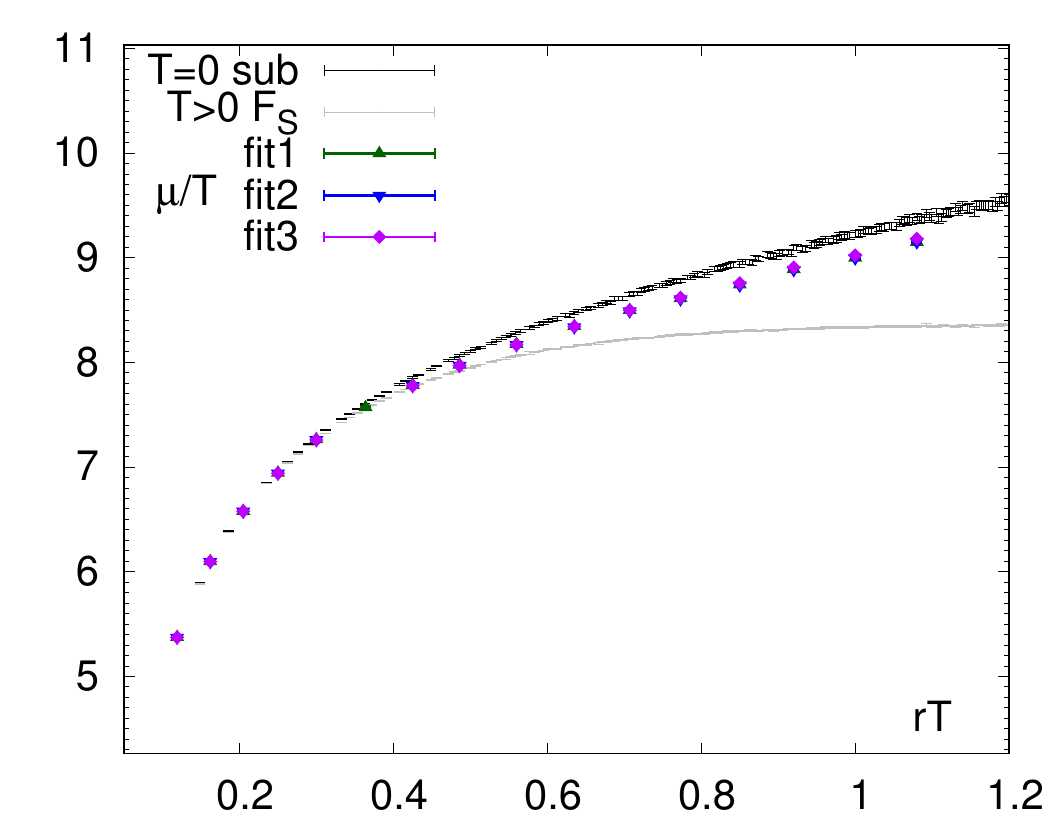}
\includegraphics[width=7.3cm]{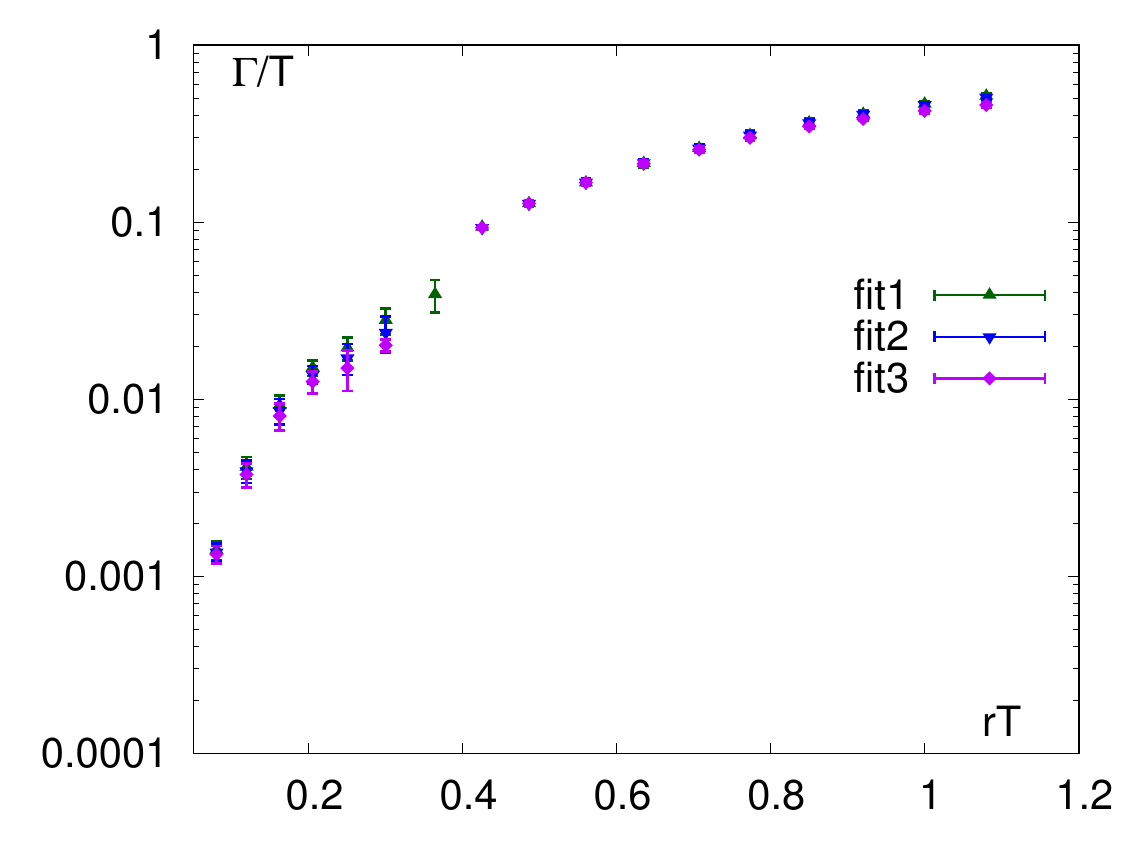}
\caption{The real (left) and the imaginary (right) parts of the potential
for $T=407$ MeV. In the left panel the zero temperature potential and the singlet free energy are also shown.}
\label{fig:pot}
\end{figure}

Performing the above fits for all values of $rT$ we get the real and imaginary parts of the potential
as functions of $rT$. The results for $T=407$ MeV are shown in Fig. \ref{fig:pot}.
We compare the real part of the potential with the corresponding zero temperature result
as well as with the singlet free energy defined as $F_S(r,T)=-T \ln W_r(\tau=1/T,T)$.
We see that at short distances the potential agrees with the singlet free energy, while at
larger distances it is significantly above the singlet free energy. While the potential at
finite temperature is always smaller than the zero temperature result we do no see any indications
for the expected screened behavior at large distances. The imaginary part of the potential on
the other hand behaves as expected: it is very small at small $rT$ and seems to saturate for $r T \simeq 1$.
However, the imaginary part is larger than in HTL perturbation theory.

\section{Conclusions}
We discussed a lattice calculation of the static $Q\bar Q$ potential
using the moments of the correlation function. We used a Lorentzian form of the spectral
function and fitted the lattice results on the moments to obtain the real and imaginary
parts of the potential. The contribution from the high energy part of the spectral
function has been subtracted using the fits of zero temperature correlators. We find
that our fit procedure works very well but
the real part of the potential is much larger than the singlet free energy and does
not show a screening behavior. We think this is due to the fact that the subtraction of the high
energy contribution from the correlator is too simplistic. In order to make our strategy for
calculating the potential at non-zero temperature viable a better modeling of the high energy
part of the spectral function will be needed. The work on this is in progress.

\noindent
{\bf Acknowledgments}\\
noindent
This work was supported by U.S. Department of Energy under Contract No. DE-SC0012704. 
We acknowledge the support by the DFG Cluster of Excellence ``Origin and Structure of the Universe''. The calculation have been carried out on the computing facilities of the Computational Center for Particle and Astrophysics (C2PAP).





\bibliographystyle{elsarticle-num}
\bibliography{ref}







\end{document}